\documentclass[prl,twocolumn]{revtex4}
\usepackage{dcolumn}
\usepackage{multirow}
\usepackage{graphicx}
\usepackage{amssymb}
\usepackage{bm}
\usepackage{hyperref}
\usepackage{epstopdf}
\usepackage{color}
\usepackage{mathrsfs}
\usepackage{amsmath,amssymb,amsthm}
\usepackage{rotating}
\usepackage{sverb, longtable}
\usepackage{subfigure}

\usepackage[graphicx]{realboxes}
\usepackage{adjustbox}

\begin{document}

\title{Primordial Gravitational Waves 2024}
\author{Deng Wang}
\email{dengwang@ific.uv.es}
\affiliation{Instituto de F\'{i}sica Corpuscular (CSIC-Universitat de Val\`{e}ncia), E-46980 Paterna, Spain}

\begin{abstract}
Primordial gravitational waves have crucial implications for the origin of the universe and fundamental physics. Using currently available cosmic microwave background data from Planck, ACT and SPT separately or their combinations with BK18 B-mode polarization and DESI observations, we give the strongest constraints on primordial gravitational waves so far.

\end{abstract}
\maketitle

{\it Introduction.} Inflation \cite{Brout:1977ix,Starobinsky:1980te,Kazanas:1980tx,Sato:1981qmu,Guth:1980zm,Linde:1981mu,Linde:1983gd,Albrecht:1982wi}, a quasi-exponential expansion phase of the very early universe after the Big Bang singularity, is believed to be responsible for the large scale homogeneity of the universe and the origin of small scale density fluctuations in the cosmic web. It not only solves the horizon and flatness problems \cite{Guth:1980zm} but also leaves imprints in the anisotropies of cosmic microwave background (CMB) observations \cite{Baumann:2009ds} and large scale matter distribution \cite{Krauss:2010ty}. However, the nature of inflation together with dark matter and dark energy is still unknown so far \cite{Martin:2013tda,Achucarro:2022qrl}. Besides generating the scalar perturbations that seed the cosmic structure, inflation can also produce the quantum tensor perturbations, i.e., the so-called primordial gravitational waves (PGWs), which carry extremely important information of the very early universe. Unlike the electromagnetic messenger CMB that only probes back to the recombination epoch, PGWs can explore the high energy physics at $\sim10^{-34}$ s very close to the start of the universe, due to feeble interaction with matter density perturbations during the travel through large cosmological distances \cite{Allahverdi:2020bys}. A successful detection of PGWs could provide the energy scale of inflation and even the first experimental demonstration of quantization of gravitational interactions \cite{Baumann:2009ds}.    

Generally, there are two main approaches to detect the PGWs, namely direct and indirect experiments. Direct detectors such as the LIGO-Virgo gravitational wave (GW) observatories where laser interferometers are used to detect the perturbations of space from astrophysical or cosmological sources, are not sensitive enough to detect PGWs at high frequencies and can just place upper bounds on the amplitude of PGWs at current stage \cite{LIGOScientific:2016jlg}. Indirect measurements are obtained by investigating the effect of PGWs on the polarization spectrum of CMB at low frequencies \cite{Planck:2018vyg,Planck:2018jri,Seljak:1996gy}. Currently, using a data combination of Planck CMB, baryon acoustic oscillations (BAO), BICEP2, Keck Array and BICEP3 CMB polarization experiments up to the 2018 observing season, the BICEP/Keck Collaboration gives the $2\,\sigma$ upper bound on the tensor-to-scalar ratio $r<0.036$ \cite{BICEP:2021xfz}, which is a clear improvement of the $2\,\sigma$ upper limit $r<0.1$ \cite{Planck:2018vyg,Planck:2018jri} from Planck alone. The combination of direct and indirect approaches can further compress the parameter space of tensor modes \cite{Planck:2018jri,Galloni:2022mok}. Nonetheless, due to the limitation of experimental sensitivity and the excess foreground contamination of astrophysical processes, it seems to be difficult to probe PGWs for direct detectors, and consequently CMB polarization data with appropriate experimental settings have a great potential to settle the issue of whether PGWs exist \cite{Seljak:1996gy}. It is worth noting that the local pulsar timing arrays (PTAs) \cite{NANOGrav:2023gor,NANOGrav:2023hvm} confront the similar problem of diverse GW sources like LIGO-Virgo or future LISA \cite{Ricciardone:2016ddg}, although they can constrain primordial tensor modes by identifying inflationary gravitational waves as the sole source.       

The next-generation CMB experiments such as CMB-S4 \cite{CMB-S4:2020lpa}, Simons Observatory \cite{Hertig:2024gxz}, LiteBIRD \cite{Matsumura:2013aja} and CORE \cite{CORE:2016ymi} aim at searching for the definite evidence of cosmic inflation and promise the detection of PGWs at the level of $\sigma(r)\sim10^{-3}$. This predicted precision may help find the imprints of PGWs, but two important questions in light of current CMB observations should be correctly answered before that: (i) Have PGWs been detected? (ii) If not, what precision of $r$ could be required to find out PGWs? For this purpose, one needs to reanalyze the available CMB data and check all the details when implementing cosmological constraints on PGWs. In doing so, via a logarithmic prior of $r$, we obtain, so far, the most stringent constraints on PGWs using currently available CMB datasets and their combinations with B-mode polarization and BAO observations.

{\it Model.} Cosmic inflation is driven by a scalar field, {\it inflaton}, with a canonical kinetic term slowly rolling in the framework of general relativity. It can be characterized by a phenomenological fluid with a significantly negative pressure \cite{Brout:1977ix,Starobinsky:1980te,Kazanas:1980tx,Sato:1981qmu,Guth:1980zm,Linde:1981mu,Linde:1983gd,Albrecht:1982wi}. During the inflation epoch, comoving tensor fluctuations are amplified from quantum vacuum fluctuations to become highly squeezed states resembling classical states. After inflation, these tensor modes reenter the Hubble horizon and evolves over time, and finally act as one kind of source of stochastic gravitational wave background observed by local GW experiments. Same as the Planck 2018 analysis \cite{Planck:2018vyg,Planck:2018jri}, we take the following primordial tensor power spectrum (PTPS) for a single-field slow rolling inflation

\begin{equation}
\mathcal{P}_{T}(k)=A_t\left(\frac{k}{k_\star}\right)^{n_t+\frac{1}{2}n_{trun}\ln\left(\frac{k}{k_\star}\right)}, \label{eq:PTS}
\end{equation}  
where $k$, $k_\star$, $A_t$, $n_t$ and $n_{trun}$ denote the comoving wavenumber, tensor pivot scale, amplitude of PTPS, tensor spectral index and running of tensor spectral index, respectively. $r\equiv A_t/A_s$, where $A_s$ is the amplitude of primordial scalar power spectrum. Throughout this work, we use the next-order inflation consistency relation adopted by the Planck collaboration, i.e., $n_t=-r(2-r/8-n_s)/8$ and $n_{trun}=r(r/8+n_s-1)/8$ \cite{Planck:2018vyg,Planck:2018jri,Planck:2015sxf}, where $n_s$ is the scalar spectral index.

{\it Data and methodology.} To probe the primordial tensor parameter space, we use the following observational datasets:

$\bullet$ CMB. CMB observations have extremely important implications for cosmology and astrophysics. They have measured the matter components, the topology and the large scale structure of the universe. We adopt the Planck 2018 high-$\ell$ \texttt{plik} temperature (TT) likelihood at multipoles $30\leqslant\ell\leqslant2508$, polarization (EE) and their cross-correlation (TE) data at $30\leqslant\ell\leqslant1996$, and the low-$\ell$ TT \texttt{Commander} and \texttt{SimAll} EE likelihoods at $2\leqslant\ell\leqslant29$ \cite{Planck:2019nip}. We employ conservatively the Planck lensing likelihood \cite{Planck:2018lbu} from \texttt{SMICA} maps at $8\leqslant\ell \leqslant400$. We also consider the Atacama Cosmology Telescope (ACT) DR4 TTTEEE data \cite{ACT:2020frw,ACT:2020gnv} at $350<\ell<8000$ and the South Pole Telescope (SPT) TTTEEE likelihood \cite{SPT-3G:2021eoc,SPT-3G:2022hvq} at $750\leqslant\ell<3000$. 

$\bullet$ BK18. To constrain the inflationary physics, we use the BICEP2, Keck Array and BICEP3 CMB polarization data up to and including the 2018 observing season \cite{BICEP:2021xfz}. This dataset, which aims at detecting the CMB B-modes, includes the additional Keck Array observations at 220 GHz and BICEP3 observations at 95 GHz relative to the previous 95/150/220 GHz dataset. We refer to this dataset as ``BK18''.

$\bullet$ BAO. BAO \cite{SDSS:2005xqv,2dFGRS:2005yhx} are rather clean probes to explore the evolution of the universe over time, which are unaffected by the nonlinear physics at small scales. Measuring the positions of these oscillations in the matter power spectrum at different redshifts can give strong constraints on the cosmic expansion history. We adopt the latest 12 DESI BAO measurements specified in \cite{DESI:2024mwx}, including the BGS sample in the redshift range $0.1 < z < 0.4$, LRG samples in $0.4 < z < 0.6$ and $0.6 < z < 0.8$, combined LRG and ELG sample in $0.8 < z < 1.1$, ELG sample in $1.1 < z < 1.6$, quasar sample in $0.8 < z < 2.1$ and the Lyman-$\alpha$ Forest sample in $1.77 < z < 4.16$ \cite{DESI:2024uvr,DESI:2024lzq}.

As previous analyses \cite{Planck:2018vyg,Planck:2018jri,BICEP:2021xfz}, the theory to be tested with data is still the $\Lambda$-cold dark matter ($\Lambda$CDM) model plus the tensor-to-scalar ratio $r$. The only difference is replacing the traditional uniform prior with logarithmic prior for $r$. This improvement can help better capture the detailed information of PGWs at the order of magnitude that is much lower than current $2\sigma$ upper bound $r\sim0.04$ \cite{BICEP:2021xfz}, and even give the realistic range of $r$.  

In order to perform the Bayesian analysis, we use the publicly available Boltzmann solver \texttt{CAMB} \cite{Lewis:1999bs} to calculate the theoretical power spectrum and employ the Monte Carlo Markov Chain (MCMC) method to infer the posterior distributions of model parameters via the online package \texttt{CosmoMC} \cite{Lewis:2002ah,Lewis:2013hha}. We analyze the MCMC chains using the public package \texttt{Getdist} \cite{Lewis:2019xzd}. The convergence rule of MCMC runs is the Gelman-Rubin diagnostic $R-1\lesssim 0.01$ \cite{Gelman:1992zz}. We adopt the following uniform priors for model parameters: the baryon fraction $\Omega_bh^2 \in [0.005, 0.1]$, cold dark matter fraction $\Omega_ch^2 \in [0.001, 0.99]$, amplitude of primordial scalar power spectrum $\mathrm{ln}(10^{10}A_s) \in [2, 4]$, scalar spectral index $n_s \in [0.8, 1.2]$, acoustic angular scale at the recombination epoch $100\theta_{MC} \in [0.5, 10]$, optical depth due to reionization $\tau \in [0.01, 0.8]$ and the logarithmic tensor-to-scalar ratio $\log_{10}r \in [-30, 5]$. The pivot scale we use is 0.05 Mpc$^{-1}$. For convenience, we refer to the data combinations of BK18+Planck+DESI, BK18+ACT+DESI and BK18+SPT+DESI as ``BPD'', ``BAD'' and ``BSD'', respectively.

\begin{figure*}
	\centering
	\includegraphics[scale=0.5]{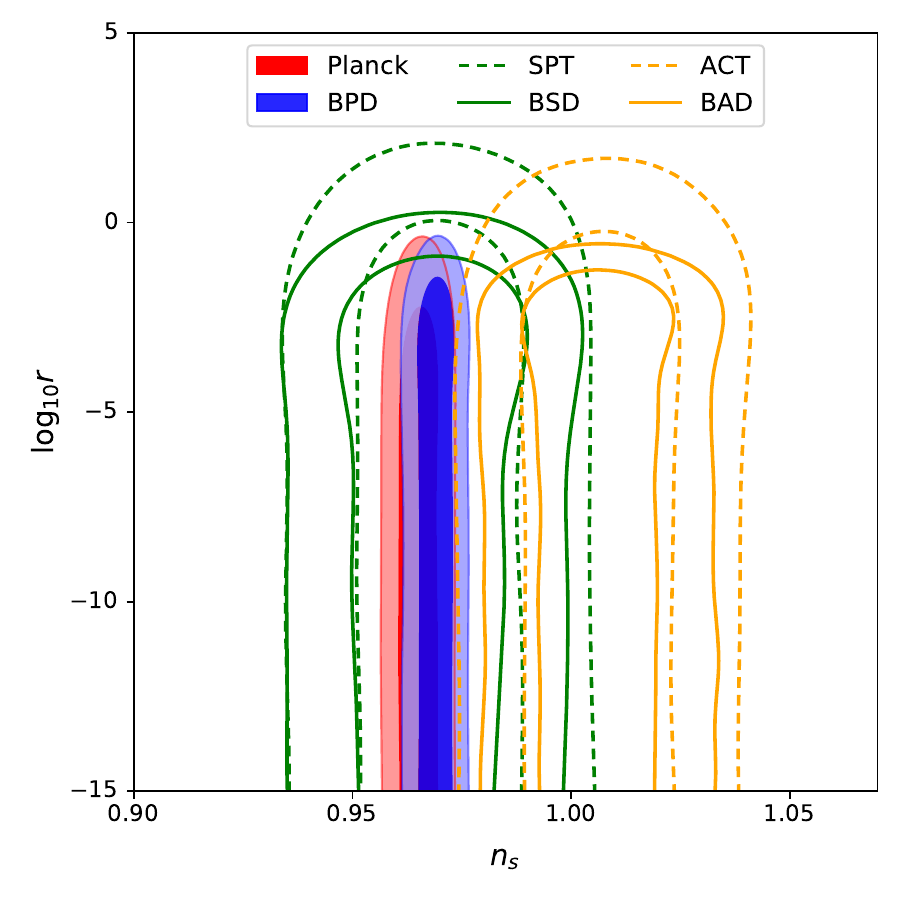}
	\includegraphics[scale=0.5]{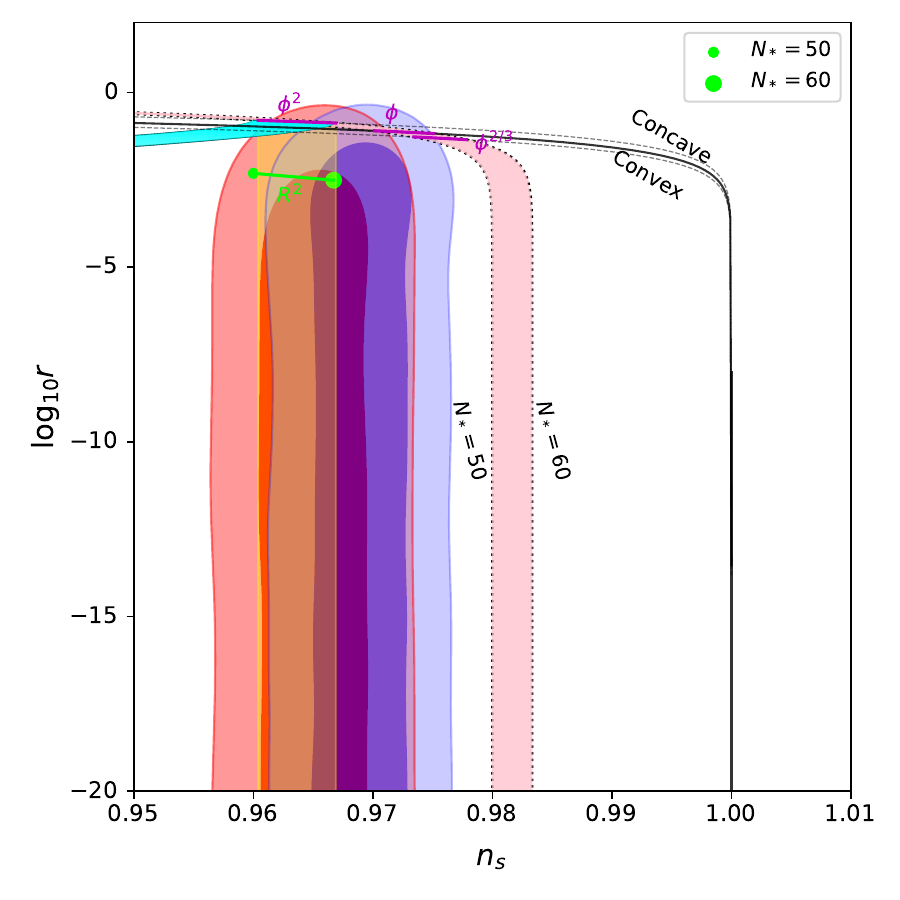}
	\caption{{\it Left.} The two-dimensional marginalized posterior distributions of the parameter pair ($n_s$, $\log_{10}r$) from the Planck (red), ACT (dashed orange), SPT (dashed green), BPD (blue), BAD (solid orange) and BSD (solid green) observations, respectively. {\it Right.} The comparison between the allowed $1\,\sigma$ and $2\,\sigma$ ($n_s$, $\log_{10}r$) contours from Plank and BPD data and the theoretical predictions of selected inflationary scenarios including $R^2$ inflation, $\alpha$ attractors, power-law inflation and natural inflation. Here $\phi$ is the inflaton field and $N_*\equiv{\rm ln}a$ ($a$ is scale factor) denotes the e-folding number.}\label{fig:pgw}
	
\end{figure*}

\begin{table*}[!t]
	\renewcommand\arraystretch{1.6}
	\begin{center}
		\caption{The $2\,\sigma$ (95\%) upper bounds on the logarithmic tensor-to-scalar ratio $\log_{10}r$ from different datasets.}
		\setlength{\tabcolsep}{2mm}{
			\label{tab:pgw}
			\begin{tabular}{c|c|c|c|c|c|c}
				\hline
				\hline
				Data & Planck & ACT & SPT & BPD & BAD & BSD  \\
				\hline 
				$\log_{10}r$ &  $<-2.433$ \,\, ($2\,\sigma$) &  $<-0.643$ \,\, ($2\,\sigma$) &  $<-0.299$ \,\, ($2\,\sigma$) &  $<-2.273$ \,\, ($2\,\sigma$) &  $<-2.178$ \,\, ($2\,\sigma$) &  $<-1.796$ \,\, ($2\,\sigma$) \\
				\hline
				\hline
		\end{tabular}}
	\end{center}
\end{table*} 

{\it Results.} In light of currently available CMB, B-modes and BAO observations, our numerical results are presented in Fig.\ref{fig:pgw} and Tab.\ref{tab:pgw}. Overall, we give the tightest upper bounds on PGWs at the $2\,\sigma$ confidence level (CL) using three CMB experiments separately and their combinations with BK18 CMB B-modes and the latest DESI BAO observations. For the CMB-only cases, Planck provides the tightest constraint $r<0.0037$ at the $2\,\sigma$ CL, which compresses the $r$ range by $\sim$ 96\% relative to 0.1 reported by the Planck collaboration \cite{Planck:2018vyg}. ACT's $2\,\sigma$ constraint $r<0.2274$ gives a clearly larger $\log_{10}r$ bound than Planck. Interestingly, the fact that ACT prefers $n_s=1$, namely the scale-invariance, still remains here. Even though another ground-based telescope SPT provides the loosest constraint $r<0.5026$, it favors the similar $n_s$ value to the satellite experiment Planck. It is noteworthy that $n_s$ constraints from all three CMB facilities are very consistent with the original results reported by each collaboration \cite{Planck:2018vyg,ACT:2020gnv,SPT-3G:2022hvq}, since tensor fluctuations are insensitive to other physical processes. Our analyses reveals that, ACT and SPT can give complementary constraints on CMB B-modes and new viewpoints, while Planck provides the strongest constraint on PGWs so far.       

Furthermore, the addition of BK18 and DESI data just slightly shrinks the parameter spaces of ($n_s$, $\log_{10}r$) for Planck, but reduces clearly the parameter spaces for ACT and SPT. This is mainly because BK18 prefers a similar upper limit of $\log_{10}r$ to Planck when assuming a fiducial cosmology and can not give a clear lower bound on $\log_{10}r$. Therefore, BPD, BAD and BSD give close $2\,\sigma$ upper limits in Fig.\ref{fig:pgw}. It is interesting to see that $1\,\sigma$ upper limit of $\log_{10}r$ from BPD is larger than the CMB-only constraint. At first glance, this seems to be anomalous based on the experience that a combined dataset should give a tighter constraint than a single dataset. Nonetheless, these results are actually reasonable and subtle because the addition of BK18 data stretches 
the $1\,\sigma$ bounds of $\log_{10}r$ to be very close to $2\,\sigma$ bounds based on the fact that the one-dimensional posterior probability density distribution of $\log_{10}r$ from the BK18 data has a sharp peak around $r\sim10^{-2}$ and no lower bound on $r$ (see Fig.\ref{fig:pgw}). This means that $1\,\sigma$ $\log_{10}r$ bound given by BPD is very close to its $2\,\sigma$ upper limit. Hence, BPD has affected the tensor parameter space compared to their corresponding CMB-only cases. Moreover, the reason why BPD prefers a larger $n_s$ than Plank alone is that DESI BAO can increase the Hubble constant $H_0$ \cite{DESI:2024mwx} when combined with the comoving sound horizon at the drag epoch from Planck \cite{Planck:2018vyg}.

Since all datasets have provided new constraints on PGWs in logarithmic space in single field slow-roll inflation, it is important and meaningful to compare current constraints with theoretical predictions of different inflation models. Specifically, we compare the predictions of $R^2$ inflation \cite{Starobinsky:1980te,Mukhanov:1981xt,Starobinsky:1982ee}, inflationary $\alpha$-attractors \cite{Kallosh:2013yoa,Kallosh:2021mnu}, power-law inflation \cite{Linde:1983gd,McAllister:2008hb,Silverstein:2008sg} and natural inflation \cite{Freese:1990rb,Adams:1992bn} with the most stringent constraint from Planck and BPD. In the right panel of Fig.\ref{fig:pgw}, considering the e-folds range $N_*\in[50,60]$, we find that natural inflation and models with concave potential are clearly ruled out. It is interesting that the power-law inflation that has a potential $V(\phi)\propto\phi^n$ (roughly $n>\frac{2}{3}$) is basically ruled out at the $2\,\sigma$ CL. Interestingly, ranging from $N_*=50$ to $N_*=60$, $R^2$ inflation which is believed to the best model by the Planck collaboration \cite{Planck:2018vyg} is now excluded around the $2\,\sigma$ CL. It is clear that there is still a large theory space needed to be explored in light of current constraints on PGWs.

{\it Discussions and conclusions.} During the almost two decades, PGWs are always constrained via the linear prior of $r$. Employing the logarithmic prior, we give the tightest constraints on PGWs at beyond the $2\,\sigma$ CL in light of three available CMB datasets and their combinations with BK18 and DESI observations. The tightest constraint we obtain for the CMB-only cases is $r<0.0037$ at the $2\,\sigma$ CL from Planck. When comparing the theoretical predictions of selected inflationary scenarios with the most stringent constraint from BPD, we find that natural inflation and models with concave potential are obviously ruled out, and power-law inflation and $R^2$ inflation are basically excluded at $2\,\sigma$ CL. Our constraints could change the research status of inflationary cosmologies and inspire new explorations of inflation theories so as to explain currently allowed parameter space by data. For example, $r=10^{-25}-10^{-3}$ is urgent to be interpreted by new physics of the very early universe. 

Interestingly, it seems that we can not conclude that the scale-invariance is completely ruled out, even if ACT's constraint on $n_s$ is inconsistent with Planck, WMAP \cite{WMAP:2012nax} and SPT. This issue should be addressed by future high precision CMB observations such as CMB-S4 \cite{CMB-S4:2020lpa}, Simons Observatory \cite{Hertig:2024gxz}, LiteBIRD \cite{Matsumura:2013aja} and CORE \cite{CORE:2016ymi}.

{\it Acknowledgements.} DW is supported by the CDEIGENT Fellowship of Consejo Superior de Investigaciones Científicas (CSIC). DW acknowledges the usage of Planck \cite{Planck:2018vyg}, ACT \cite{ACT:2020gnv}, SPT \cite{SPT-3G:2022hvq}, BK18 \cite{BICEP:2021xfz} and DESI \cite{DESI:2024mwx} data.


\begin{thebibliography}{99}

\bibitem{Brout:1977ix}
R.~Brout, F.~Englert and E.~Gunzig,
``The Creation of the Universe as a Quantum Phenomenon,'' 
Annals Phys. \textbf{115}, 78 (1978).

\bibitem{Starobinsky:1980te}
A.~A.~Starobinsky,
``A New Type of Isotropic Cosmological Models Without Singularity,''
Phys. Lett. B \textbf{91}, 99-102 (1980).

\bibitem{Kazanas:1980tx}
D.~Kazanas,
``Dynamics of the Universe and Spontaneous Symmetry Breaking,''
Astrophys. J. Lett. \textbf{241}, L59-L63 (1980).

\bibitem{Sato:1981qmu}
K.~Sato,
``First-order phase transition of a vacuum and the expansion of the Universe,''
Mon. Not. Roy. Astron. Soc. \textbf{195}, no.3, 467-479 (1981).

\bibitem{Guth:1980zm}
A.~H.~Guth,
``The Inflationary Universe: A Possible Solution to the Horizon and Flatness Problems,''
Phys. Rev. D \textbf{23}, 347-356 (1981).

\bibitem{Linde:1981mu}
A.~D.~Linde,
``A New Inflationary Universe Scenario: A Possible Solution of the Horizon, Flatness, Homogeneity, Isotropy and Primordial Monopole Problems,''
Phys. Lett. B \textbf{108}, 389-393 (1982).

\bibitem{Linde:1983gd}
A.~D.~Linde,
``Chaotic Inflation,''
Phys. Lett. B \textbf{129}, 177-181 (1983).

\bibitem{Albrecht:1982wi}
A.~Albrecht and P.~J.~Steinhardt,
``Cosmology for Grand Unified Theories with Radiatively Induced Symmetry Breaking,''
Phys. Rev. Lett. \textbf{48}, 1220-1223 (1982).

\bibitem{Baumann:2009ds}
D.~Baumann,
``Inflation,''
[arXiv:0907.5424 [hep-th]].

\bibitem{Krauss:2010ty}
L.~Krauss, S.~Dodelson and S.~Meyer,
``Primordial Gravitational Waves and Cosmology,''
Science \textbf{328} (2010), 989-992.


\bibitem{Martin:2013tda}
J.~Martin, C.~Ringeval and V.~Vennin,
``Encyclop\ae{}dia Inflationaris,''
Phys. Dark Univ. \textbf{5-6} (2014), 75-235.

\bibitem{Achucarro:2022qrl}
A.~Ach\'ucarro \textit{et al.},
``Inflation: Theory and Observations,''
[arXiv:2203.08128 [astro-ph.CO]].

\bibitem{Allahverdi:2020bys}
R.~Allahverdi \textit{et al.},
``The First Three Seconds: a Review of Possible Expansion Histories of the Early Universe,''
[arXiv:2006.16182 [astro-ph.CO]].

\bibitem{LIGOScientific:2016jlg}
B.~P.~Abbott \textit{et al.} [LIGO-Virgo Collaboration],
``Upper Limits on the Stochastic Gravitational-Wave Background from Advanced LIGO\textquoteright{}s First Observing Run,''
Phys. Rev. Lett. \textbf{118}, no.12, 121101 (2017).

\bibitem{Planck:2018vyg}
N.~Aghanim \textit{et al.} [Planck Collaboration],
``Planck 2018 results. VI. Cosmological parameters,''
Astron. Astrophys. \textbf{641}, A6 (2020)
[erratum: Astron. Astrophys. \textbf{652}, C4 (2021)].

\bibitem{Planck:2018jri}
Y.~Akrami \textit{et al.} [Planck Collaboration],
``Planck 2018 results. X. Constraints on inflation,''
Astron. Astrophys. \textbf{641}, A10 (2020).

\bibitem{Seljak:1996gy}
U.~Seljak and M.~Zaldarriaga,
``Signature of gravity waves in polarization of the microwave background,''
Phys. Rev. Lett. \textbf{78}, 2054-2057 (1997).

\bibitem{BICEP:2021xfz}
P.~A.~R.~Ade \textit{et al.} [BICEP/Keck Collaboration],
``Improved Constraints on Primordial Gravitational Waves using Planck, WMAP, and BICEP/Keck Observations through the 2018 Observing Season,''
Phys. Rev. Lett. \textbf{127}, no.15, 151301 (2021).

\bibitem{Galloni:2022mok}
G.~Galloni \textit{et al.},
``Updated constraints on amplitude and tilt of the tensor primordial spectrum,''
JCAP \textbf{04}, 062 (2023).

\bibitem{NANOGrav:2023gor}
G.~Agazie \textit{et al.} [NANOGrav Collaboration],
``The NANOGrav 15 yr Data Set: Evidence for a Gravitational-wave Background,''
Astrophys. J. Lett. \textbf{951}, no.1, L8 (2023).

\bibitem{NANOGrav:2023hvm}
A.~Afzal \textit{et al.} [NANOGrav Collaboration],
``The NANOGrav 15 yr Data Set: Search for Signals from New Physics,''
Astrophys. J. Lett. \textbf{951}, no.1, L11 (2023).

\bibitem{Ricciardone:2016ddg}
A.~Ricciardone,
``Primordial Gravitational Waves with LISA,''
J. Phys. Conf. Ser. \textbf{840}, no.1, 012030 (2017).

\bibitem{CMB-S4:2020lpa}
K.~Abazajian \textit{et al.} [CMB-S4 Collaboration],
``CMB-S4: Forecasting Constraints on Primordial Gravitational Waves,''
Astrophys. J. \textbf{926}, no.1, 54 (2022).

\bibitem{Hertig:2024gxz}
E.~Hertig \textit{et al.} [SO Collaboration],
``The Simons Observatory: Combining cross-spectral foreground cleaning with multi-tracer $B$-mode delensing for improved constraints on inflation,''
[arXiv:2405.01621 [astro-ph.CO]].

\bibitem{Matsumura:2013aja}
T.~Matsumura \textit{et al.} [LiteBIRD Collaboration],
``Mission design of LiteBIRD,''
J. Low Temp. Phys. \textbf{176}, 733 (2014).

\bibitem{CORE:2016ymi}
F.~Finelli \textit{et al.} [CORE Collaboration],
``Exploring cosmic origins with CORE: Inflation,''
JCAP \textbf{04}, 016 (2018)

\bibitem{Planck:2015sxf}
P.~A.~R.~Ade \textit{et al.} [Planck Collaboration],
``Planck 2015 results. XX. Constraints on inflation,''
Astron. Astrophys. \textbf{594}, A20 (2016).

\bibitem{Planck:2019nip}
N.~Aghanim \textit{et al.} [Planck Collaboration],
``Planck 2018 results. V. CMB power spectra and likelihoods,''
Astron. Astrophys. \textbf{641}, A5 (2020).

\bibitem{Planck:2018lbu}
N.~Aghanim \textit{et al.} [Planck Collaboration],
``Planck 2018 results. VIII. Gravitational lensing,''
Astron. Astrophys. \textbf{641}, A8 (2020).


\bibitem{ACT:2020frw}
S.~K.~Choi \textit{et al.} [ACT Collaboration],
``The Atacama Cosmology Telescope: a measurement of the Cosmic Microwave Background power spectra at 98 and 150 GHz,''
JCAP \textbf{12}, 045 (2020).

\bibitem{ACT:2020gnv}
S.~Aiola \textit{et al.} [ACT Collaboration],
``The Atacama Cosmology Telescope: DR4 Maps and Cosmological Parameters,''
JCAP \textbf{12}, 047 (2020).

\bibitem{SPT-3G:2021eoc}
D.~Dutcher \textit{et al.} [SPT Collaboration],
``Measurements of the E-mode polarization and temperature-E-mode correlation of the CMB from SPT-3G 2018 data,''
Phys. Rev. D \textbf{104}, no.2, 022003 (2021).

\bibitem{SPT-3G:2022hvq}
L.~Balkenhol \textit{et al.} [SPT Collaboration],
``Measurement of the CMB temperature power spectrum and constraints on cosmology from the SPT-3G 2018 TT, TE, and EE dataset,''
Phys. Rev. D \textbf{108}, no.2, 023510 (2023).

\bibitem{SDSS:2005xqv}
D.~J.~Eisenstein \textit{et al.} [SDSS Collaboration],
``Detection of the Baryon Acoustic Peak in the Large-Scale Correlation Function of SDSS Luminous Red Galaxies,''
Astrophys. J. \textbf{633}, 560-574 (2005).

\bibitem{2dFGRS:2005yhx}
S.~Cole \textit{et al.} [2dFGRS Collaboration],
``The 2dF Galaxy Redshift Survey: Power-spectrum analysis of the final dataset and cosmological implications,''
Mon. Not. Roy. Astron. Soc. \textbf{362}, 505-534 (2005).

\bibitem{DESI:2024mwx}
A.~G.~Adame \textit{et al.} [DESI Collaboration],
``DESI 2024 VI: Cosmological Constraints from the Measurements of Baryon Acoustic Oscillations,''
[arXiv:2404.03002 [astro-ph.CO]].

\bibitem{DESI:2024uvr}
A.~G.~Adame \textit{et al.} [DESI Collaboration],
``DESI 2024 III: Baryon Acoustic Oscillations from Galaxies and Quasars,''
[arXiv:2404.03000 [astro-ph.CO]].

\bibitem{DESI:2024lzq}
A.~G.~Adame \textit{et al.} [DESI Collaboration],
``DESI 2024 IV: Baryon Acoustic Oscillations from the Lyman Alpha Forest,''
[arXiv:2404.03001 [astro-ph.CO]].

\bibitem{Lewis:1999bs}
A.~Lewis, A.~Challinor and A.~Lasenby,
``Efficient computation of CMB anisotropies in closed FRW models,''
Astrophys. J. \textbf{538}, 473-476 (2000).

\bibitem{Lewis:2002ah}
A.~Lewis and S.~Bridle,
``Cosmological parameters from CMB and other data: A Monte Carlo approach,''
Phys. Rev. D \textbf{66}, 103511 (2002).

\bibitem{Lewis:2013hha}
A.~Lewis,
``Efficient sampling of fast and slow cosmological parameters,''
Phys. Rev. D \textbf{87}, no.10, 103529 (2013).

\bibitem{Lewis:2019xzd}
A.~Lewis,
``GetDist: a Python package for analysing Monte Carlo samples,''
[arXiv:1910.13970 [astro-ph.IM]].

\bibitem{Gelman:1992zz}
A.~Gelman and D.~B.~Rubin,
``Inference from Iterative Simulation Using Multiple Sequences,''
Statist. Sci. \textbf{7}, 457-472 (1992).

\bibitem{Mukhanov:1981xt}
V.~F.~Mukhanov and G.~V.~Chibisov,
``Quantum Fluctuations and a Nonsingular Universe,''
JETP Lett. \textbf{33}, 532-535 (1981).

\bibitem{Starobinsky:1982ee}
A.~A.~Starobinsky,
``Dynamics of Phase Transition in the New Inflationary Universe Scenario and Generation of Perturbations,''
Phys. Lett. B \textbf{117}, 175-178 (1982).

\bibitem{Kallosh:2013yoa}
R.~Kallosh, A.~Linde and D.~Roest,
``Superconformal Inflationary $\alpha$-Attractors,''
JHEP \textbf{11}, 198 (2013).

\bibitem{Kallosh:2021mnu}
R.~Kallosh and A.~Linde,
``BICEP/Keck and cosmological attractors,''
JCAP \textbf{12}, no.12, 008 (2021).

\bibitem{McAllister:2008hb}
L.~McAllister, E.~Silverstein and A.~Westphal,
``Gravity Waves and Linear Inflation from Axion Monodromy,''
Phys. Rev. D \textbf{82}, 046003 (2010).

\bibitem{Silverstein:2008sg}
E.~Silverstein and A.~Westphal,
``Monodromy in the CMB: Gravity Waves and String Inflation,''
Phys. Rev. D \textbf{78}, 106003 (2008).

\bibitem{Freese:1990rb}
K.~Freese, J.~A.~Frieman and A.~V.~Olinto,
``Natural inflation with pseudo - Nambu-Goldstone bosons,''
Phys. Rev. Lett. \textbf{65}, 3233-3236 (1990).

\bibitem{Adams:1992bn}
F.~C.~Adams, J.~R.~Bond, K.~Freese, J.~A.~Frieman and A.~V.~Olinto,
``Natural inflation: Particle physics models, power law spectra for large scale structure, and constraints from COBE,''
Phys. Rev. D \textbf{47}, 426-455 (1993).

\bibitem{WMAP:2012nax}
G.~Hinshaw \textit{et al.} [WMAP Collaboration],
``Nine-Year Wilkinson Microwave Anisotropy Probe (WMAP) Observations: Cosmological Parameter Results,''
Astrophys. J. Suppl. \textbf{208}, 19 (2013).



















\end{thebibliography}
\end{document}